# Terahertz and Far-Infrared Windows Opened at Dome A, Antarctica

Sheng-Cai Shi[1]*, Scott Paine[2], Qi-Jun Yao[1], Zhen-Hui Lin[1], Xin-Xing Li[1†], Wen-Ying Duan[1], Hiroshi Matsuo[3], Qizhou Zhang[2], Ji Yang[1], M.C.B. Ashley[4], Zhaohui Shang[5,7], Zhong-Wen Hu[6]

[1]Purple Mountain Observatory, Key Laboratory of Radio Astronomy, Chinese Academy of Sciences, Nanjing, China.

[2]Smithsonian Astrophysical Observatory, Cambridge, MA, USA.

[3]National Astronomical Observatory of Japan, Mitaka, Tokyo, Japan.

[4]The University of New South Wales, Sydney, Australia.

[5]Tianjin Normal University, Tianjin, China.

[6]Nanjing Institute of Astronomical Optics and Technology, National Astronomical Observatories, Chinese Academy of Sciences, Nanjing, China.

[7]National Astronomical Observatories, Chinese Academy of Sciences, Beijing, China

†Suzhou Institute of Nano-Tech and Nano-Bionics, Chinese Academy of Sciences, Suzhou, China

*Correspondence to: scshi@pmo.ac.cn.

**The terahertz and far-infrared (FIR) band, from approximately 0.3 THz to 15 THz (1 mm to 20 µm), is important for astrophysics as the thermal radiation of much of the universe peaks at these wavelengths and many spectral lines that trace the cycle of interstellar matter also lie within this band[1-8]. However, water vapor renders the terrestrial atmosphere opaque to this frequency band over nearly all of the Earth's surface[9]. Early radiometric measurements[10] below 1 THz at Dome A (80°22′ S, 77°21′ E), the highest point of the cold and dry Antarctic ice sheet, suggest that it may offer the best possible access for ground-based astronomical observations in the terahertz and FIR band. To address uncertainty in radiative transfer modelling, we carried out measurements of atmospheric radiation from Dome A spanning the entire water vapor pure rotation band from 20 µm to 350 µm wavelength by a Fourier transform spectrometer (FTS). Our measurements expose atmospheric windows having significant transmission throughout this band. Furthermore, by combining our broadband spectra with auxiliary data on the atmospheric state over Dome A, we set new constraints on the spectral absorption of water vapor at upper tropospheric temperatures important for accurately modeling the terrestrial climate. In particular, we find that current spectral models significantly underestimate the $H_2O$ continuum absorption.**

We directly measured the zenith atmospheric radiation at Dome A throughout the $H_2O$ pure rotation band, i.e. the terahertz and far-infrared (FIR) band. The motivation for this work is China's ongoing development of Dome A as a scientific base including facilities for astronomy[10, 11]. Such broadband measurements from this site have not been made before and are needed because narrowband radiometry combined with modeling is insufficient, primarily due to limited understanding of the $H_2O$ continuum absorption, especially at the low temperatures typical of Dome A. Ground-based measurements are needed because satellite measurements have inadequate sensitivity in the lower troposphere. Our results were obtained with a remotely operated Fourier transform spectrometer (FTS)[12], which was deployed to Dome A by the 26th Chinese Antarctic Research Expedition (CHINARE) team during the 2009 / 2010 traverse. The FTS ran for 19 months until August 2011, measuring the down-welling zenith sky radiance from 0.75 THz – 15 THz with 13.8 GHz spectral resolution. Technical details regarding the instrument and calibration methods are given in Methods. As discussed there, to achieve sufficient signal-to-noise ratio spectra were averaged into 6-hour bins, and good radiance calibration was achieved at frequencies above 1 THz.

For the site survey statistics, we derived transmittance spectra from the measured radiance spectra using an isothermal radiative transfer approximation as described in Methods. The fractional accuracy of the transmittance thus obtained corresponds to the fractional change in the absolute atmospheric temperature across the scale height of the water vapor column, which is approximately 5%. Transmittance errors associated with strong water lines terminated within the instrument or in the surface boundary layer can be recognized by their correspondence to the well-known positions of these lines. The key advantage of the isothermal transmittance approximation, in contrast with the more detailed spectroscopic study below, is that it requires no assumptions regarding the vertical structure of the atmosphere.

The transmittance statistics are presented in Fig. 1, where we show quartile transmittance spectra for the entire year (a, c), and for winter (April – September) only (b, d). To correct for uneven sampling over the calendar year, when compiling percentile statistics we applied a de-biasing procedure whereby spectra obtained in calendar periods that were sampled only once during the two years of operation were duplicated, and brief gaps in sampling were covered by interpolating between temporally adjacent spectra. Sufficiently high transmittance (~20%) to support astronomical observation is consistently observed in winter in the three windows at 1.03, 1.3 and 1.5 THz encompassing the astrophysically important spectral lines indicated in Fig. 1. Much higher (>40%) far-infrared transmittance is found in a number of windows starting at 7.1 THz. We note that although our spectra are limited to frequencies above 1 THz, they are consistent with Dome A being an excellent year-round observing site in the submillimeter windows below 1 THz.

Given plans for future observatories and scientific facilities at Dome A, the exposed terahertz and FIR windows present a unique opportunity for ground-based astronomy. For example, unique terahertz spectral line transitions from atomic and molecular species, such as C+ at 1.90 THz (158 μm), N+ at 1.46 THz (205 μm), $H_2D$+ at 1.37 THz

(219 μm), and high-J CO lines, can be observed to trace the life cycle of stars and interstellar matter as well as chains of chemical reactions that ultimately shape the chemical makeup of planetary systems like our own. In addition, spectral lines of species such as $O^{2+}$ at frequencies greater than 7 GHz allow exploration of the energy balance in the interstellar medium. As a ground-based site, Dome A can support larger facilities, either single telescopes with large aperture or multiple telescopes phased up as an interferometer, with more rapid and agile development cycles than space or aircraft-based platforms.

Our measurements also allow us to address an issue in atmospheric science relating to uncertainties in the absorption spectrum of water vapor, associated with poorly understood collisional effects which give rise to smoothly-varying continuum absorption[13, 14] that has significant impact on atmospheric radiation models[15-18]. The importance of measuring the continuum absorption has motivated several recent field experiments, including the RHUBC campaigns[19-21], ECOWAR[22, 23], and CAVIAR[24]. However, the full atmospheric temperature range has not been adequately explored. The measurements we report here from Dome A provide new constraints on the $H_2O$ continuum associated with heterogeneous collisions (known as the foreign continuum) towards the low range of atmospheric temperature. In addition, the low $H_2O$ column density over Dome A gives unprecedented access to wavelengths in the core of the $H_2O$ rotation band where previous measurements have lacked sensitivity.

We studied the water vapor absorption spectrum by combining our broadband spectra with auxiliary data on the atmospheric state over Dome A. Given the atmospheric state in the form of vertical profiles of temperature and water vapor concentration over the site, a radiative transfer model can be used to predict the observed radiance spectrum and thereby test the spectral absorption data used in the model. With the exception of the $H_2O$ foreign continuum absorption, the spectral absorption data used in the radiative transfer model are well constrained by laboratory measurements. Therefore, the residuals between the measured and modeled spectra can be interpreted in terms of an implied correction to the continuum absorption coefficient. The radiative transfer model employed the MT_CKD (v. 2.5.2) water vapor continuum[14]. We chose MT_CKD as the reference continuum model for this study because it has been extensively validated in laboratory and field experiments, albeit at higher temperatures than we access here, and because it is widely used in atmospheric radiation codes incorporated into climate models.

Our starting point for estimating the atmospheric state is the NASA MERRA reanalysis[25] interpolated over Dome A. MERRA provides an estimate of the global atmospheric state constrained by satellite, surface, and upper air measurements with 6-hour resolution matching the temporal averaging bins of our spectra. The ranges of MERRA-derived vertical temperature and water vapor profiles during the period of this study (August 2010) are shown in Fig. 2. The modest interquartile variation observed in these profiles together with relatively high data quality in our spectra is the reason we chose to focus on this period. Recent dropsonde validations[26] of the satellite-derived temperature profiles over Antarctica give us high confidence in the data constraining the MERRA temperature profiles. Moreover, as expected the MERRA water vapor profile in the lower tropopause closely tracks the ice saturation vapor pressure profile. As indicated in Fig. 2, the range of temperature across the water vapor column was small during the study period.

MERRA does not resolve the strong winter Antarctic surface inversion[27], and the absolute accuracy of the MERRA water vapor profile is uncertain. Therefore we used an analysis method in which we divided the measured spectral interval into two parts – the band-edge frequencies (f < 3.6 THz and f > 12.5 THz) comprising spectral channels where the observed radiance is insensitive to the continuum, and the mid-band frequencies (3.6 THz < f < 12.5 THz) where the radiance in the transmission windows includes a significant continuum contribution. The band edge channels were used with a radiative transfer model[28] to fit a two-parameter adjustment to the initial MERRA-derived profiles. These two parameters were a scaling factor on the MERRA water vapor profile in the troposphere, and the base temperature of a surface layer modeling the surface inversion.

An example of one of these fits is shown in Fig. 3a. With the atmospheric state anchored to the band edge channels, we find that the radiances in the mid-band windows are consistently underestimated. Using the channel-by-channel derivative of the model radiance with respect to the foreign-continuum absorption coefficient, the residuals in the windows were used to derive an implied adjustment to the absorption coefficient. Fig. 3b compares the MT_CKD absorption coefficient with the quartile statistics of the adjusted values found for all of the spectra included in this study, covering a range of water vapor column density from 70 µm to 220 µm PWV. The consistent value found for the adjusted foreign continuum absorption coefficient across this wide range of water vapor column density, particularly in the higher signal-to-noise windows from 5 THz – 9 THz, is strong evidence that the residuals are indeed associated with water vapor absorption as opposed to systematic calibration error, errors in the dry air spectroscopy, or absorption by other atmospheric constituents such as hydrometeors. The smooth trend in the adjusted continuum, as well as the generally good fit near the centers of unsaturated water lines, would also appear to rule out errors in the $H_2O$ line-by-line spectroscopy as the origin of the radiance residuals. Note that the implied adjustment to the MT_CKD foreign continuum absorption coefficient, at the column density weighted mean $H_2O$ temperature of 218 K over these spectra, is as high as a factor of 2.5 in the mid-band windows just below 9 THz.

By reaching new extremes of low temperature and accessing windows in the core of the $H_2O$ pure rotation band, our measurements at Dome A have provided new constraints on the spectral absorption of water vapor important for modeling radiative processes in the cold upper troposphere and for retrieving atmospheric properties from outbound spectral radiance measurements. Taken together with the transmittance statistics for the astronomical observing windows discussed above, our measurements have demonstrated the value of this unique site to both astronomy and atmospheric science.

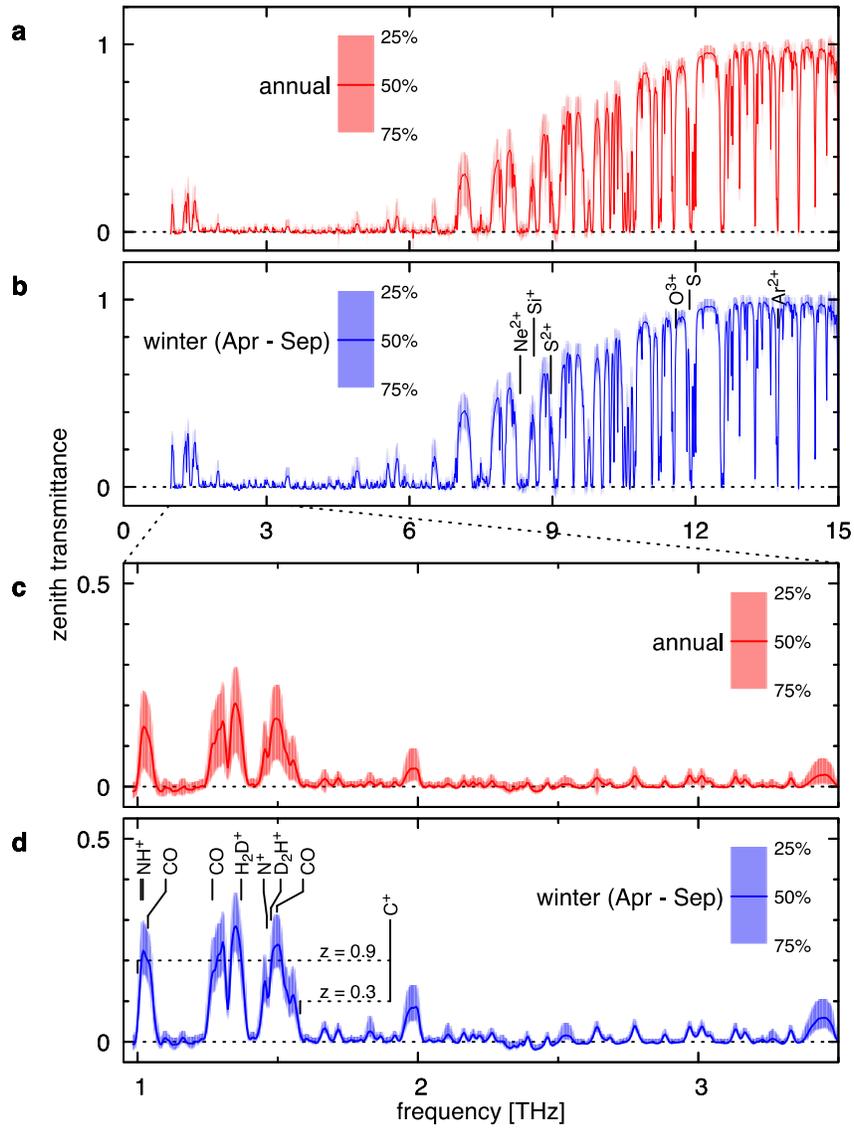

**Figure 1:** Annual and winter (April-September) transmittance spectra measured at Dome A during 2010-11. Quartile statistics for each frequency channel were compiled independently. Heavy lines indicate median transmittances, and the shaded region shows the interquartile range. To account for irregular time sampling throughout the year, data were de-biased as described in the text. Frequencies of several astrophysically-important spectral lines are indicated; note that some of these lines are observable only at certain non-zero redshift $z$.

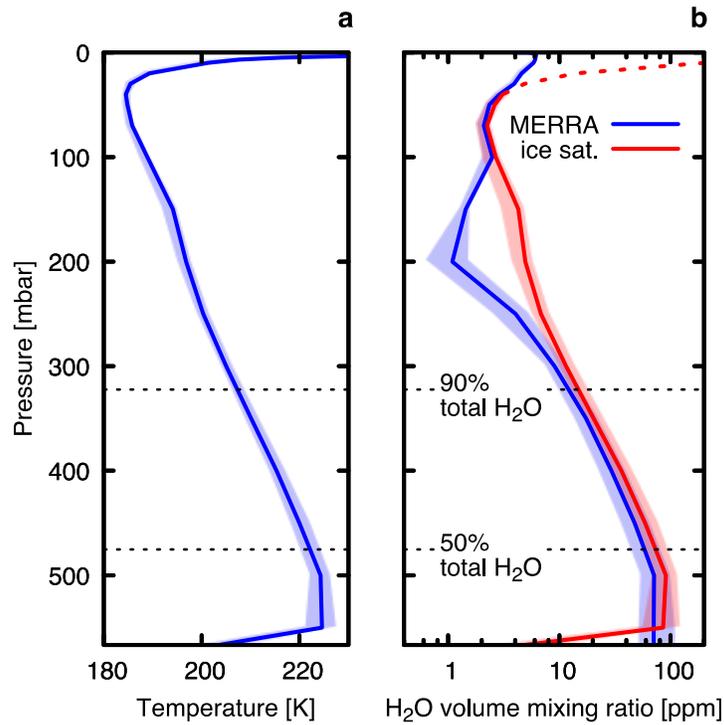

**Figure 2:** MERRA vertical profiles of temperature (**a**), and $H_2O$ mixing ratio (**b**), during the August 2010 study period. The temperature point at the surface, associated with the strong winter surface inversion, is from our own instrument. Shading indicates the interquartile range of variation. For comparison with the MERRA water vapor profile, the red profile on the right shows the saturated mixing ratio over ice corresponding to the temperature profile on the left. Horizontal dotted lines indicate pressure altitudes below which the median MERRA $H_2O$ profile contains 90% and 50% respectively of the total $H_2O$ column density, indicating the small range of temperature across most of the $H_2O$ column. The column density weighted mean $H_2O$ temperature for the study period was 218 K.

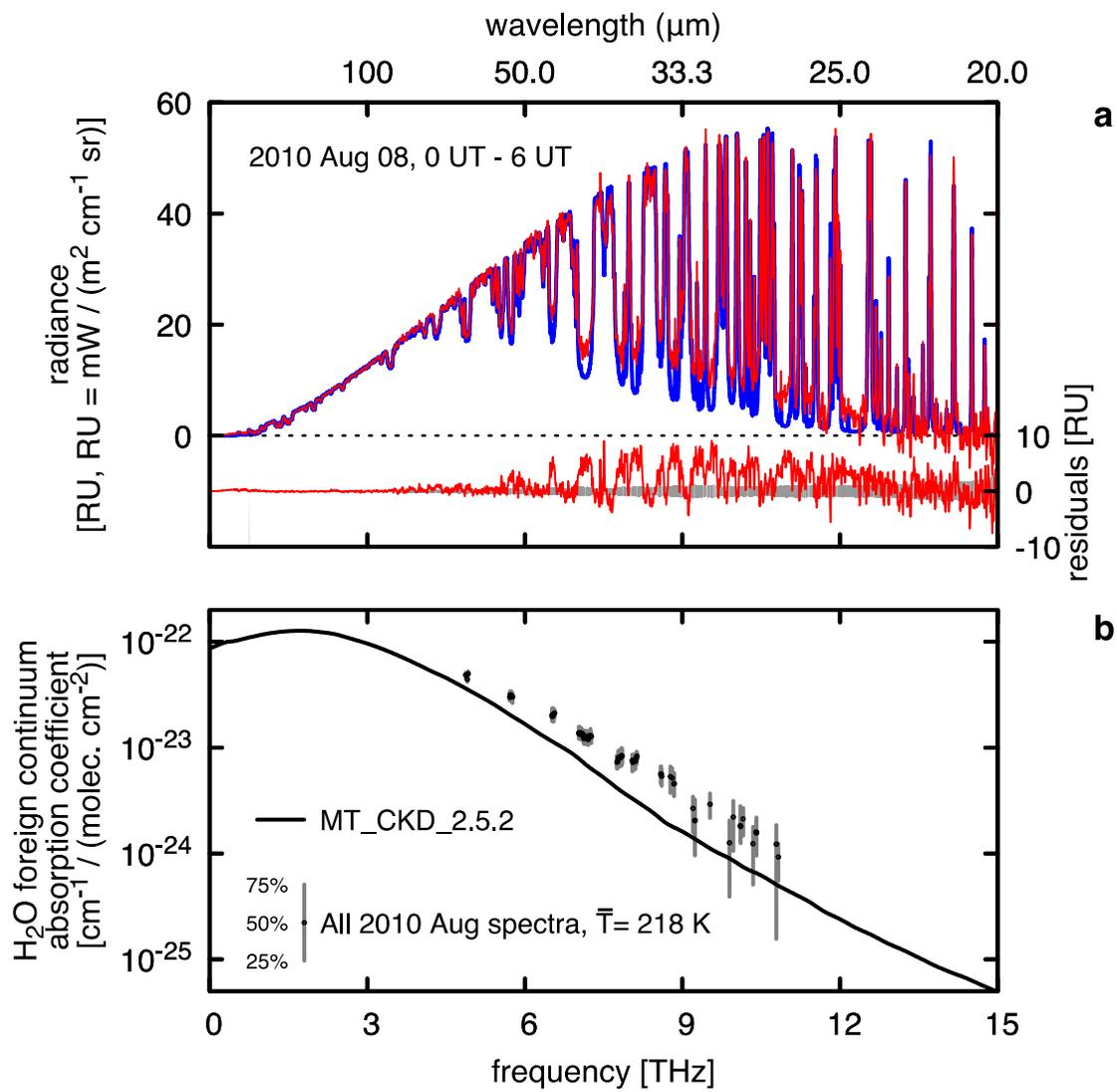

**Figure 3**: (**a**) An example of a spectral fit from the August 2010 data set. The measured radiance spectrum is shown in red, and the model spectrum computed from the scaled MERRA profiles is shown in blue. Note the significant residuals in the spectral windows within the $H_2O$ rotation band; the gray band plotted along with the residuals indicates the noise level in the measured spectrum. (**b**) Quartile statistics (symbols) of the adjusted $H_2O$ foreign continuum absorption coefficient derived from the radiance residuals for all August 2010 spectra compared with MT_CKD v. 2.5.2 (solid line). The vertical scale is shown in the units customarily used in the literature, with the radiation term $\nu \tanh(h\nu/kT)$ factored out. Channels used in this analysis were screened using criteria described in Methods.

**Methods**

Description of the Instrument

The Dome A FTS is a polarizing Michelson (Martin-Puplett) interferometer[29]. The conceptual and optical designs were developed at PMO and SAO, and are described in Li, et al.[12] Detailed engineering and fabrication of the interferometer, detectors, and band-defining filters were carried out by Blue Sky Spectroscopy (Lethbridge, Canada), and QMC Instruments (Cardiff, UK). Photos of the instrument and its installation at Dome A in the UNSW PLATO support module[30] are presented in Extended Data Fig. 1.

At Dome A the instrument was required to run unattended on low power, placing significant constraints on detector sensitivity and calibration. Use of the cryogenically cooled low-noise detectors typically employed in similar instruments to achieve background-limited sensitivity was not feasible, and instead the instrument used DLATGS pyroelectric detectors operated at ambient temperature. Actively cooled or heated blackbody loads for calibration were also not practical within the instrument power budget. Instead, calibration was accomplished using two passive loads. These were an internal warm reference load terminating one of the interferometer input ports, and an outdoor cold calibration load on the roof of PLATO, which could swing under computer control to intercept the view towards the zenith sky. Extended Data Fig. 2 plots the time history of these two load temperatures over the course of the campaign. The typical temperature difference available for calibration was approximately 40 K in summer and 50 K in winter. The loads were made using TKRAM, a product of Thomas Keating Limited (Billingshurst, UK) having less than 1% power reflectivity at THz frequencies.

The spectral coverage of the FTS, 750 GHz – 15 THz, was split into two bands defined by different low-pass filters in front of the detector at each of the two output ports of the interferometer. The high band admitted all power up to 15 THz, while the low band was restricted to frequencies below 3.6 THz. This allowed the low band detector to operate at increased optical throughput, and also reduced radiative background noise due to the steeply increasing Planck radiance and associated atmospheric brightness fluctuations towards higher frequencies. The 750 GHz low frequency coverage limit of the FTS was determined by the signal-to-noise ratio achieved in the low band given the sensitivity of the FTS combined with the temporal stability of the atmosphere. The 15 THz high frequency limit was selected to substantially cover the $H_2O$ pure rotation band, extending high enough in frequency to include several optically-thin windows and unsaturated $H_2O$ emission lines.

Spectral resolution was set by the interferometer scan length and weighting applied for apodization. The scan length was ±12.5 mm relative to the zero path delay (ZPD) position of the interferometer, and a Blackman weighting function was applied in processing. This yielded an instrumental resolution function with a half-width of 13.8 GHz, chosen to be narrow enough to isolate a number of unsaturated $H_2O$ lines, but otherwise as broad as possible in order to increase optical throughput and channel bandwidth given the other constraints on sensitivity noted above.

Spectral Calibration and averaging

At Dome A, each scan of the interferometer took 25 s. Interferograms were stored as 8 integrated scans, alternating after each integration between viewing the zenith sky or the external passive calibration load. A complete measurement cycle comprising one integration on the sky and one on the calibration load thus took approximately 9 minutes. Each sky/load pair of interferograms (for each of the two detector bands) was Fourier transformed and processed to yield calibrated spectra using a complex-domain calibration method similar to that described by Revercomb et al.[31] Below, we describe the calibration of a single detector band. In processing, the spectra for the two bands were combined in a channel-by-channel noise-weighted average to produce a single merged spectrum for each sky/load measurement cycle.

An important factor in the calibration is that both the sky and the outside calibration load were viewed through a 2 mm thick high-density polyethylene (HDPE) window at the top of a tube passing through the insulated roof of the PLATO module, so that the window absorption and reflection need to be accounted for. The window, located directly below the calibration load, was tilted to avoid interference effects and to direct the reflected view onto absorbing material lining the tube. The bottom of the tube was covered with a thin polyethylene membrane having negligible optical reflection and absorption serving to prevent convective heat transfer to the interior of PLATO, so that the window and upper end of the tube would follow the outdoor temperature. We therefore assume the window absorption and reflection are terminated at the outside calibration load temperature.

Under this assumption, during the calibration phase when the two input ports of the interferometer view the reference load and calibration load, the raw spectrum obtained by complex Fourier transformation of the interferogram can be written as

$$S_{cal}(\nu) = R_c(\nu) \cdot \left(B_\nu(T_{cc}) - B_\nu(T_{rc})\right) \qquad (1)$$

Here, $R_c$ is the complex responsivity of the FTS during the calibration phase, $B_\nu(T_{cc})$ is the Planck radiance at the temperature $T_{cc}$ of the calibration load during the calibration phase of the measurement cycle, and $B_\nu(T_{rc})$ is the Planck radiance at the temperature $T_{rc}$ of the reference load during the calibration cycle.

During the on-sky phase of the measurement cycle the raw spectrum is

$$S_{sky}(\nu) = R_s(\nu) \cdot \left[\mathcal{T}_w(\nu)I_{sky}(\nu) + \left(1 - \mathcal{T}_w(\nu)\right)B_\nu(T_{cs}) - B_\nu(T_{rs})\right] \qquad (2)$$

where

$$\mathcal{T}_w(\nu) = (1-r)^2[1 - a(\nu)] \qquad (3)$$

is the net window transmittance approximated as the product of two surface reflections at a frequency-independent power reflectivity $r$, and a frequency-dependent single-pass absorption $a(\nu)$. $I_{sky}(\nu)$ is the downwelling zenith radiance, and $B_\nu(T_{cs})$ and $B_\nu(T_{rs})$ are the Planck radiances at the calibration load and reference load temperatures, respectively, during the sky viewing phase of the measurement cycle.

We expect the complex responsivities $R_c$ and $R_s$ to be identical apart from a linear phase term associated with any drift $\delta_{ZPD}$ in the ZPD position of the interferometer. That is,

$$R_s(\nu) = R_c(\nu) e^{2\pi i (\delta_{ZPD} \cdot \nu / c)} \qquad (4)$$

If we assume $\delta_{ZPD} = 0$, then the complex responsivities cancel in forming the ratio $S_{sky}(\nu)/S_{cal}(\nu)$, and the calibration equation for the FTS can be written as

$$I_{diff}(\nu) = I_{sky}(\nu) - I_{cc}(\nu) = \left(\frac{S_{sky}(\nu)}{S_{cal}(\nu)} - 1\right) \cdot I_0(\nu) - \Delta I(\nu) \qquad (5)$$

where

$$I_0(\nu) = \frac{1}{\mathcal{T}_w(\nu)} (B_\nu(T_{cc}) - B_\nu(T_{rc})) \qquad (6)$$

is the (typically negative) difference in load radiances, adjusted for the window transmittance, and

$$\Delta I(\nu) = \frac{1 - \mathcal{T}_w(\nu)}{\mathcal{T}_w(\nu)} (B_\nu(T_{cs}) - B_\nu(T_{cc})) - \frac{1}{\mathcal{T}_w(\nu)} (B_\nu(T_{rs}) - B_\nu(T_{rc})) \qquad (7)$$

is a much smaller term associated with changes in the load temperatures between the calibration and sky phases of a measurement cycle. Note that the effect of forming the ratio of the complex spectra $S_{sky}(\nu)$ and $S_{cal}(\nu)$ is to cancel out the phase in the complex responsivity, so that the real part $\Re\left(I_{diff}(\nu)\right)$ consists of the calibrated radiance difference between the zenith sky and cal load plus a noise component, and the imaginary part $\Im\left(I_{diff}(\nu)\right)$ consists of noise only. To estimate the noise variance in each channel of the spectrum $\Re\left(I_{diff}(\nu)\right)$, we apply a bidirectional exponential smooth to $\left|\Im\left(I_{diff}(\nu)\right)\right|^2$, with a $1/e$ width of ±5 channels. This noise variance is used to optimally weight the channel-by-channel combination of the two spectral bands of the FTS into the single merged spectrum for each measurement cycle, and to optimally weight the subsequent averages of multiple spectra into 6-hour integrations.

The effect of a small ZPD drift $\delta_{ZPD} \ll \lambda$ is a small rotation in the complex plane of the ratio spectrum $S_{sky}(\nu)/S_{cal}(\nu)$, projecting part of the real signal component onto the imaginary noise component. This would produce an undesirable bias in our noise estimates. To avoid this, we compute an estimate of $\delta_{ZPD}$ for each frequency channel in the highest signal-to-noise ratio part of the high-band ratio spectrum, from 10 THz to 14 THz, average these into a single estimate for $\delta_{ZPD}$ for the spectrum, and apply a corresponding phase correction to the ratio $S_{sky}(\nu)/S_{cal}(\nu)$ for both the high and low band spectra. The cumulative distribution (Extended Data Fig. 3) of these estimates over all sky/cal cycles in the campaign shows that $\delta_{ZPD}$ was typically less than 1 µm, much smaller than the shortest wavelength of operation $\lambda_{min} = 20$ µm. Ideally, $\delta_{ZPD}$ should have zero median if it is associated with bounded temperature variations and indeed the actual median is $\tilde{\delta}_{ZPD} = -0.03$ µm, which is negligible.

In this instrument, several factors can potentially contribute to gain errors that multiply $I_{diff}(\nu)$. These include possible thermometry error associated with temperature gradients across the calibration load between temperature sensors and the absorbing surface, and accumulation or removal of ice or snow particles on the HDPE window between sky and cal load viewing phases of the measurement cycle. The presence of several optically thin windows above 12 THz affords an opportunity to make a differential correction for gain errors under the dry conditions that prevail at this site. This correction takes advantage of the approximately linear relationship under optically thin conditions between the average radiance $I_1$ in a wide transmission window at $\bar{\nu}_1 = 12.25\ THz$, and the average radiance $I_2$ in three narrower windows with average frequency $\bar{\nu}_2 = 13.93\ THz$, which is

$$I_2 = a \cdot (I_1 - I_2) + b. \tag{8}$$

Given an initial calibration, Eq. 8 can be used to produce a refined estimate of the location of the high-frequency zero-radiance baseline based on the difference between the two near-baseline radiances $I_1$ and $I_2$. Because these radiances are already small, the coefficients $a$ and $b$ need not be known to high accuracy. Starting with median climatological profiles based on MERRA, we varied a scale factor on the tropospheric water vapor profile to find the slope $a = 1.53$, and dry intercept $b = 0.31\ RU$. Given an initial calibrated radiance $I_{diff}(\bar{\nu}_2)$, the gain correction factor $\eta$ multiplying $I_{diff}(\nu)$ that corrects the high frequency spectral baseline is

$$\eta = \frac{I_{diff}(\bar{\nu}_2) - I_2 + b}{I_{diff}(\bar{\nu}_2) - a \cdot (I_1 - I_2)}. \tag{9}$$

The gain correction factor $\eta$ was computed and applied to the 6-hour averaging bins. In approximately 11% of cases where conditions were not sufficiently dry or stable, no correction was made. For the corrected cases, the cumulative distribution of $\eta$ plotted in Extended Data Fig. 4 shows that while there was negligible systematic bias in the gain calibration during the August 2010 study period, there was a systematic gain bias of approximately 5% over the entire campaign, possibly due to solar heating of the calibration load during months when the sun is up.

The transmittance $\mathcal{T}_w(\nu)$ of the 2 mm HDPE window was calibrated in-situ using an iterative method. This was necessary because the absorption spectrum of HDPE has significant temperature dependence[32,33], and could thus be expected to change between normal room temperature and deployment to Dome A. We made an initial estimate of the single-pass absorption $a(\nu)$ based on low-temperature HDPE absorption spectra in the literature[32-34], scaling the loss tangent spectrum (i.e. the ratio of the imaginary to real refractive index) to minimize the residuals an initial model fit to the radiances in high-frequency optically-thin windows, where the radiances approach the zero radiance baseline under dry conditions. We then checked and adjusted this initial absorption estimate by performing a linear regression on the isothermally-approximated optical depth in the individual transmittance windows within the $H_2O$ rotation band against precipitable water constrained by the unsaturated $H_2O$ lines above the rotation band envelope. This analysis used the August 2010 subset of the data, which, as mentioned in the main text, had the highest quality and covered the driest period. An example of one such regression is shown in Extended Data Fig. 5. The dry intercept of the optical depth is small, and in contrast with the wet absorption it can be modeled accurately using spectral data (principally for $N_2$ collision-induced absorption) that are well constrained by laboratory measurements at the relevant temperatures[28]. For each atmospheric window within the $H_2O$ rotation band, we computed an adjustment factor on the HDPE absorption based on the difference between the modeled dry optical depth and the regression intercept.

The adjustment factors were linearly interpolated in frequency between the windows when applied to the HDPE absorption spectrum, and the FTS spectra were then recalibrated. This process was iterated to convergence, which was substantially reached after the first iteration. The initial and final estimates of $a(\nu)$ are plotted in Extended Data Fig. 6. In contrast with the absorption, the real part $n = 1.53$ of the refractive index of HDPE is essentially independent of frequency and temperature over our range of interest, so the accuracy of the reflection loss $(1 - r)^2 = 0.914$ is not a source of significant uncertainty.

We note that the analysis used to calibrate the HDPE window provided an approximate indication of an underestimated $H_2O$ continuum absorption coefficient, via comparison of the regression slope with an estimate from a model based on the monthly average atmospheric state from MERRA with a scaled water vapor profile as shown in Extended Data Fig. S5. There, we have also shown the effect of turning off the $H_2O$ continuum in the model, illustrating the relative contributions of the $H_2O$ continuum and line absorption in the 7.1 THz (42 μm) spectral window. In principle, either the line wing absorption or the continuum absorption could be increased to match the regression slope in any particular window. However, the consistency of the continuum adjustment across multiple windows bordered by $H_2O$ lines with various strengths and frequency separations strongly supports our conclusion that it is the continuum which is underestimated.

Transmittance computation and statistics

As discussed in the main text, the transmittance statistics presented in Fig. 1 were compiled from transmittance spectra derived from the measured zenith radiance spectra using an isothermal atmosphere approximation. This is defined as

$$t_{iso}(\nu) = \frac{B_\nu(T_{atm}) - I_{sky}(\nu)}{B_\nu(T_{atm}) - B_\nu(T_{cb})} \quad (10)$$

where $T_{atm}$ is the effective temperature of the atmosphere, $B_\nu(T_{atm})$ is the corresponding Planck radiance. $B_\nu(T_{cb})$ is the Planck radiance corresponding to the cosmic microwave background at $T_{cb} = 2.7$ K, which is insignificant except at our lowest measured frequencies. The celestial diffuse infrared background at higher frequencies is negligible relative to the atmospheric radiance.

As discussed in the main text, the fractional accuracy of the isothermal atmosphere approximation for transmittance over Dome A is expected to be approximately 5%, provided an appropriate value is chosen for $T_{atm}$. Surface air temperature is not a good proxy for $T_{atm}$, particularly in the Antarctic winter when surface inversion depths exceeding 20 K are often observed[27]. Instead, we derive $T_{atm}$ using the opaque channels forming the Planck baseline of the measured radiance spectrum. These channels are terminated over a range of altitude extending well above any surface inversion that may be present. The baseline is found by converting observed zenith radiance $I_{sky}(\nu)$ in each channel to the equivalent Planck brightness temperature via the inverse Planck function

$$T_b(\nu) = \frac{h\nu}{k \ln\left(1 + \frac{2h\nu^3}{c^2 I(\nu)}\right)} \quad (11)$$

We then form a histogram, weighted by signal-to-noise ratio, of the $T_b$ values for all the channels in a spectrum, in bins 1 K wide. The temperature at the center of the bin with the highest weighted count is then taken as the effective $T_{atm}$. This procedure is illustrated for an example spectrum in Extended Data Fig. 7.

Continuum Adjustment Statistics

When compiling the statistics of the water vapor continuum absorption coefficient adjustments plotted in Fig. 3 of the main text, we applied several screening criteria to select suitable spectral channels. Only those channels between 3.6 THz and 12.6 THz, which are not used in the fits that constrain the water vapor profile scaling, are included in the analysis. Within this frequency range, channels were further screened as described below.

We screened channels for zenith transmittance (based on the fitted model spectrum), requiring transmittance between $t = 0.3$ and $t = 0.9$. The purpose of this screen is to eliminate channels for which the optical depth is too high or too low. Channels with high optical depth are sensitive to the low-level structure of the water vapor and temperature profiles, potentially introducing systematic bias. Channels for which the optical depth is too low have insufficient sensitivity to the absorption coefficient relative to calibration error affecting the radiance baseline. We additionally require that the change in transmittance across the width of a single spectral channel (13.8 GHz) is $\Delta t \leq 0.05$, which serves to reject channels in the near wings of spectral lines.

For channels that pass the zenith transmittance screening, we then apply the following statistical screens. First, to ensure meaningful quartile statistics, a channel must have passed the transmittance screen in at least 16 spectra to be included in the analysis. Next, we identify clear spectral windows by requiring at least 5 adjacent channels to pass this screen. Finally, if the median value of the continuum absorption coefficient adjustment factor differs by more than 5% between adjacent channels, these channels are rejected. This last criterion serves to reject channels that may be affected by errors in the strength or broadening coefficient of spectral lines bordering the spectral transmission windows, or for which the signal-to-noise ratio is insufficient to produce an median estimate consistent to better than 5%, noting that the underlying variation in the continuum across the width of a single spectral channel will be much smaller than this.

As a test of the sensitivity of our analysis to the initial MERRA profiles, we repeated our analysis after applying ±2 K perturbations to the MERRA temperature profiles, and after replacing the MERRA water vapor profile with the ice saturation profile in the troposphere, finding that these changes did not significantly affect our results within the noise level of our spectra. This is because our scaling of the MERRA water vapor profile effectively calibrates the mid-band continuum against the band edge line absorption. As discussed above (cf. Extended Data Fig. 4), the discrepancy in the continuum absorption seen here is large enough to be seen in a simpler analysis starting with a monthly mean climatological profile. The main reason for preferring the individual 6-hourly MERRA profiles as our starting point is to minimize model bias. Our fitted $H_2O$ profile scale factors showed good agreement with the MERRA total $H_2O$ column density – the median scale factor was 1.21, and the interquartile range was 1.06 to 1.43. We note that having calibrated the MERRA reanalysis with our direct radiometric measurements, the reanalysis can be used as a tool to extend the site characterization over a longer time period stretching back to 1979.

**Acknowledgments:**

The authors gratefully acknowledge the generous assistance of the 26th and 27th CHINARE teams supported by Polar Research Institute of China and Chinese Arctic and Antarctic Administration, University of New South Wales PLATO team, CAS Center for Antarctic Astronomy team, and the other teams contributing to the operation of the Dome A facilities, in particular J.W.V. Storey, D.M. Luong-Van, Anna Moore, Carl Pennypacker, Don York, Lifang Wang, Longlong Feng, Zhenxi Zhu, Huigen Yang, Xiangqun Cui, Xiangyan Yuan, Xuefei Gong, Xu Zhou, Xiang Liu, Zhong Wang and Jiasheng Huang. The exemplary work of David Naylor and Brad Gom of Blue Sky Spectroscopy Inc. and Ken Wood of QMC Instruments Inc. is gratefully acknowledged. It is a pleasure to acknowledge valuable discussions with D. D. Turner of NOAA on complex-domain calibration of the FTS spectra, and E. J. Mlawer of AER, Inc. on the MT_CKD water vapor continuum model and its implementation in the radiative transfer code used in this work. MERRA data used in this study have been provided by the Global Modeling and Assimilation Office (GMAO) at NASA Goddard Space Flight Center through the NASA GES DISC online archive. Inquiries regarding data used in this work should be directed to the lead author.

Primary support for this research was provided by the Operation, Maintenance and Upgrading Fund for Astronomical Telescopes and Facility Instruments, budgeted from the Ministry of Finance of China (MOF) and administrated by the Chinese Academy of Sciences (CAS). The traverse team was financially supported by the Chinese Polar Environment Comprehensive Investigation & Assessment Program. The PLATO team was funded by the Australian Research Council and the Australian Antarctic Division. Iridium communications were provided by the US National Science Foundation and the US Antarctic Program. Co-authors Paine and Zhang received additional support for this work from Smithsonian Institution Endowment funds and the Smithsonian Competitive Grants Program for Science. Co-author Matsuo was supported partly by a visiting professorship of CAS for senior international scientists.

**Extended Data**

a                                                                                                                                                          b

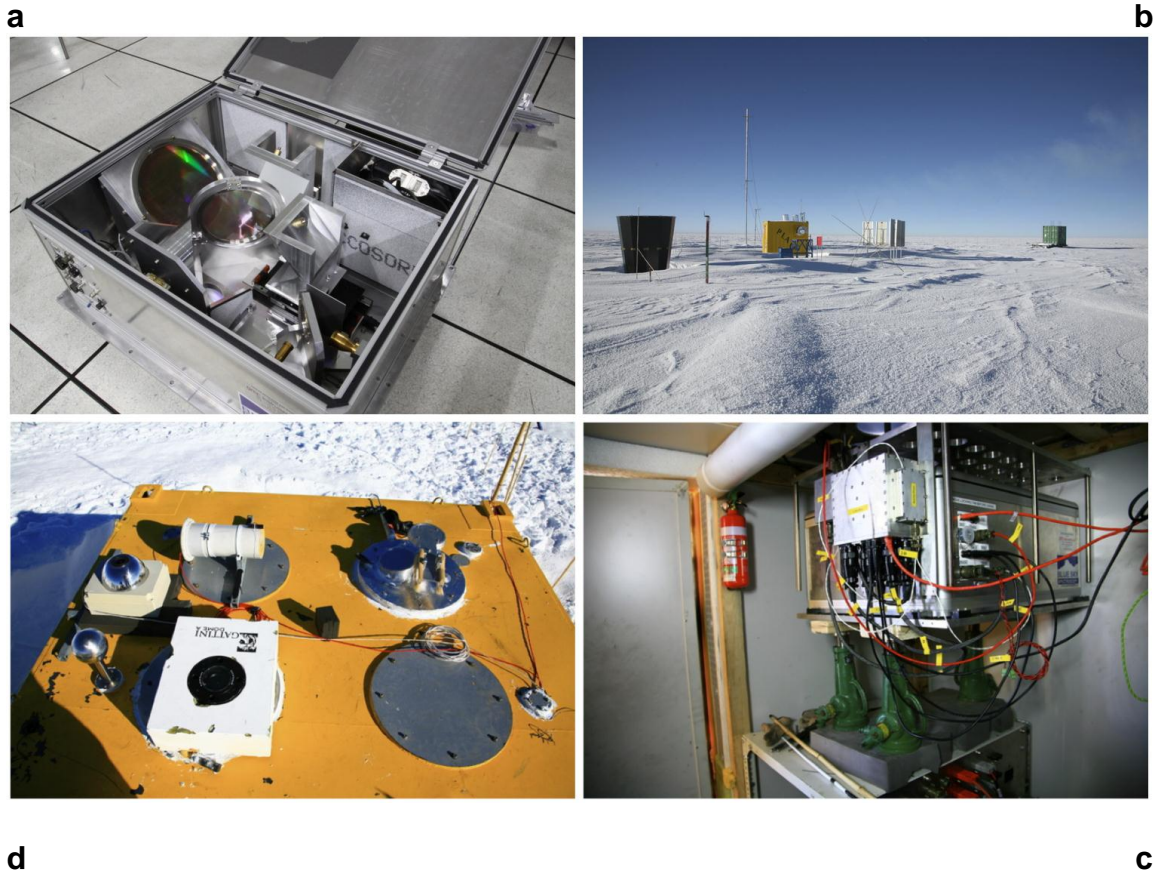

d                                                                                                                                                          c

**Extended Data Figure 1:** Photos of the FTS instrument and installation at Dome A. Clockwise from upper left: (**a**) Interior view of the FTS, with the input polarizer and internal (warm) reference load behind it at upper left, the interferometer with polarizing beam splitter and roof mirrors at center, and the output polarizer and two detectors at lower right. (**b**) View of the Dome A site showing the UNSW PLATO instrument support module. (**c**) The FTS installed in PLATO. (**d**) The roof of the PLATO module with the FTS window, outside (cold) calibration load, and window brush assembly all on the circular plate at upper right.

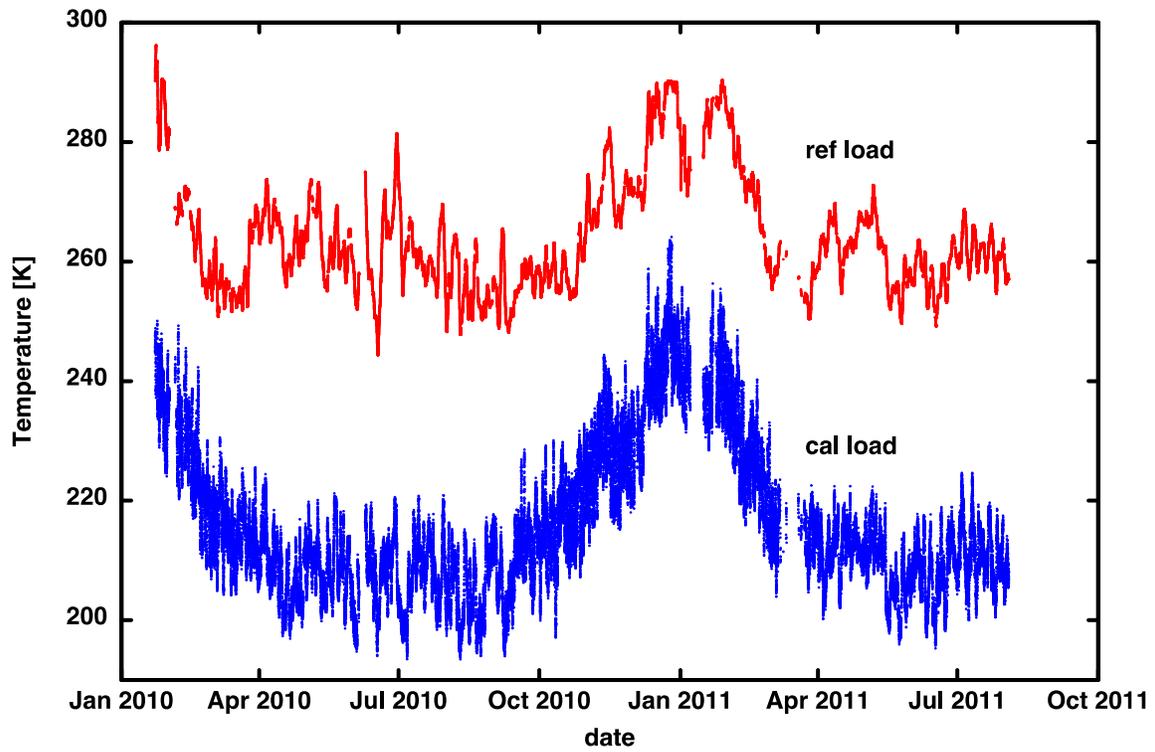

**Extended Data Figure 2:** Time series of internal reference load and outdoor calibration load temperatures over the duration of the campaign.

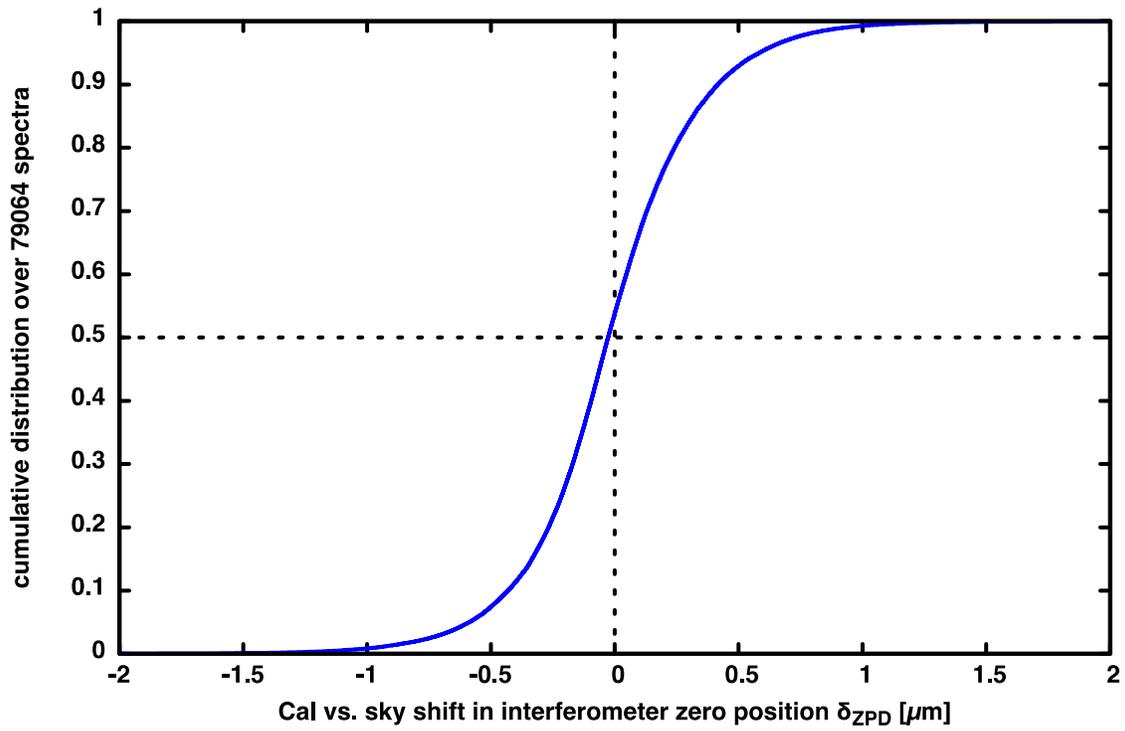

**Extended Data Figure 3:** Cumulative distribution of the estimates for $\delta_{ZPD}$, the shift in the zero path delay position of the FTS interferometer, between calibration and sky integrations over the full set of spectra acquired during the campaign.

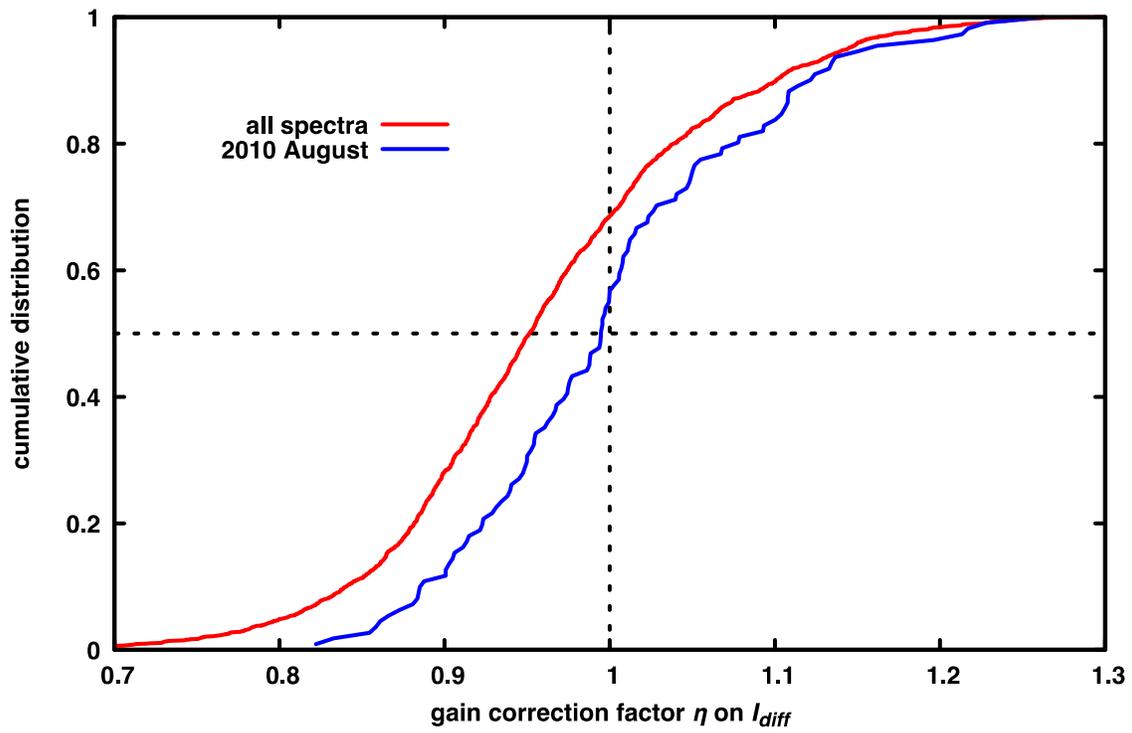

**Extended Data Figure 4:** Cumulative distributions of the baseline-based gain correction factor $\eta$ applied to $I_{diff}$ for the entire campaign (red) and for the August 2010 study period (blue).

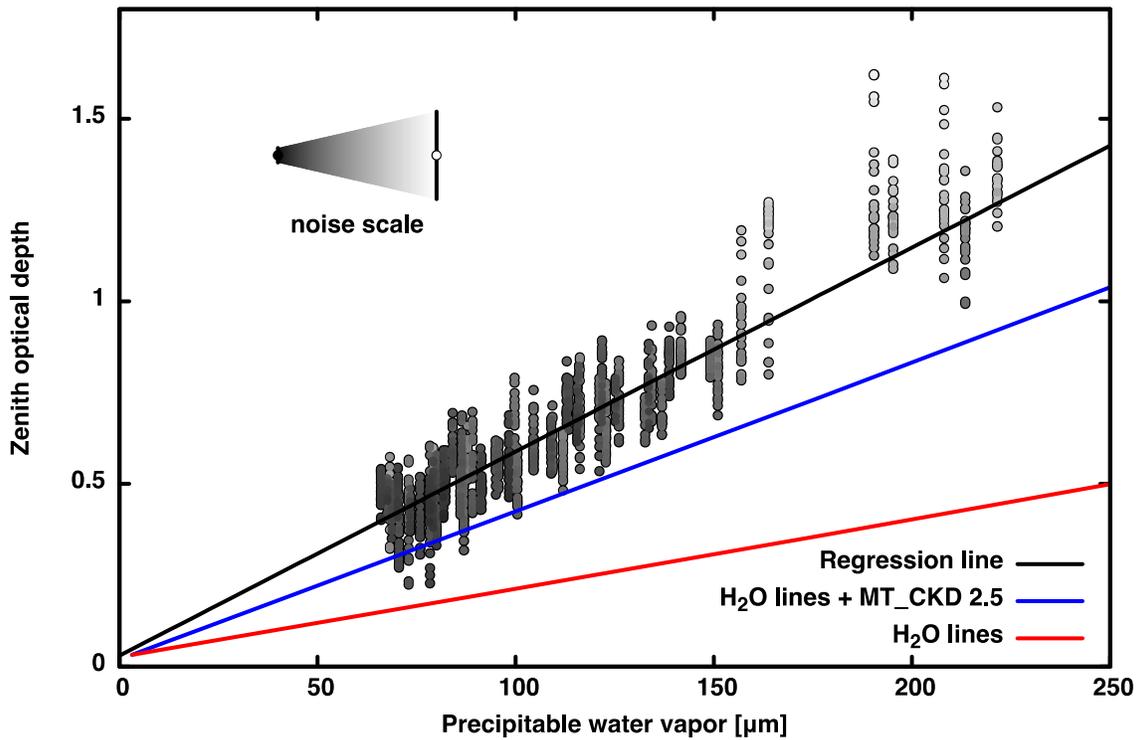

**Extended Data Figure 5:** Example of a linear regression of isothermal optical depth τ against precipitable water (PWV), in this case in the 7.1 THz (42 μm) atmospheric window during August 2010. The dry intercept, which is small and can be accurately modeled, is used to check and adjust the absorption in the HDPE window on the PLATO roof. For comparison, the expected τ vs. PWV is shown for a model based on the monthly median MERRA profiles, scaling the tropospheric water vapor profile to vary the PWV. The blue line is for the complete model using the MT_CKD v. 2.5.2 water vapor continuum. The red line is for the model with the continuum contribution turned off, to show the relative contributions from the $H_2O$ line and continuum absorption.

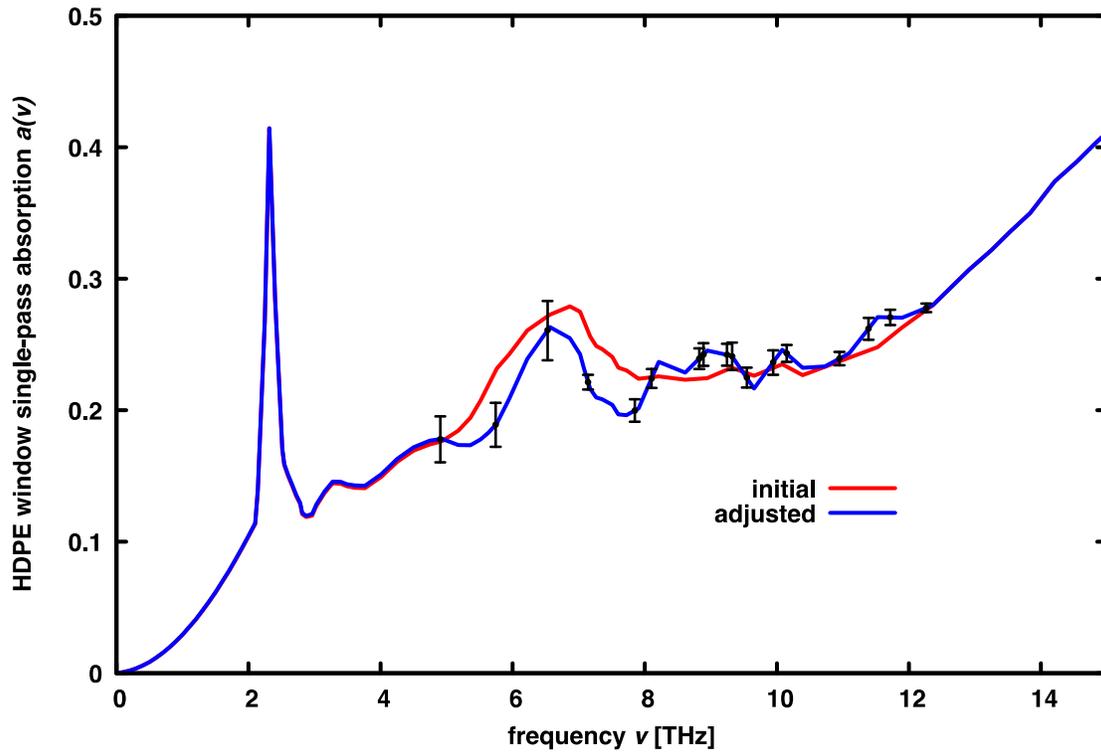

**Extended Data Figure 6:** Single-pass absorption $a(\nu)$ of the 2 mm thick HDPE window. The red curve is the initial estimated absorption spectrum, and the blue curve is the adjusted spectrum after the iterative calibration process described in the text. Points and error bars indicate window frequencies where adjustments were derived and corresponding uncertainties.

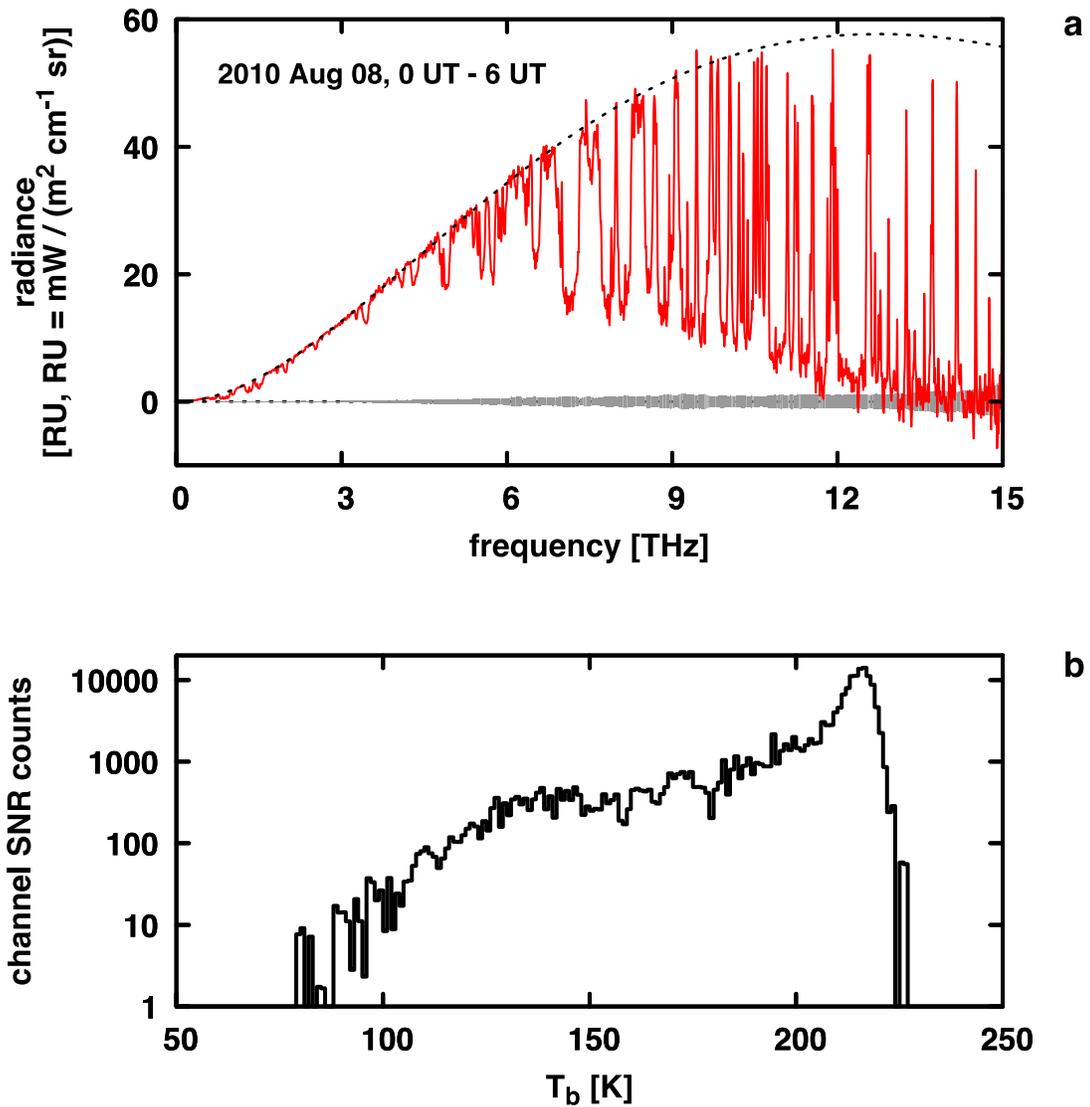

**Extended Data Figure 7:** Example of Planck baseline determination for isothermal transmittance estimates. For each 6-hour integration such as the spectrum in (a), a histogram of channel counts (b) weighted by signal-to-noise ratio is accumulated in 1 K bins of brightness temperature $T_b$ corresponding to the channel radiances. The temperature at the center of the $T_b$ bin with the highest count is used as the effective atmospheric temperature $T_{atm}$ for computing the isothermal transmittance. In this example, the peak of the histogram (b) gives $T_{atm} = 216.5$ K; the dotted curve in (a) is the corresponding Planck radiance curve $B_\nu(T_{atm})$.